\documentclass[preprint2]{aastex}

\slugcomment{}

\shorttitle{Long period X-ray source from Thorne-{\.Z}ytkow Object}
\shortauthors{Liu et al.}

\begin{document}

\title{The extremely long period X-ray source in a young supernova remnant: a Thorne-{\.Z}ytkow Object descendant?}

\author{X. W. Liu$^{1,2}$, R. X. Xu$^{2}$, E. P. J. van den Heuvel$^{3}$, G. J. Qiao$^{2}$, \\ J. L. Han$^{4}$, Z. W. Han$^{5}$, and X. D. Li$^{6}$\\
$^{1}$School of Physics and Electronic Information, China West Normal University, Nanchong 637002, China; xiongwliu@163.com\\
$^{2}$School of Physics and State Key Laboratory of Nuclear Physics and Technology, Peking University, Beijing 100871, China\\
$^{3}$Astronomical Institute Anton Pannekoek, University of Amsterdam, The Netherlands\\
$^{4}$National Astronomical Observatories, Chinese Academy of Sciences, Beijing 100012, China\\
$^{5}$Key Laboratory of the Structure and Evolution of Celestial Objects, Yunnan Observatory, Kunming 650011, China\\
$^{6}$Department of Astronomy, Nanjing University, Nanjing 210093, China}

\begin{abstract}
The origin of the 6.67 hr period X-ray source, 1E161348-5055, in the young supernova remnant RCW 103 is puzzling. We propose that it may be the descendant of a Thorne-{\.Z}ytkow Object (T{\.Z}O). A T{\.Z}O may at its formation have a rapidly spinning neutron star as a core, and a slowly rotating envelope. We found that the core could be braked quickly to an extremely long spin period by the coupling between its magnetic field and the envelope, and that the envelope could be disrupted by some powerful bursts or exhausted via stellar wind. If the envelope is disrupted after the core has spun down, the core will become an extremely long-period compact object, with a slow proper motion speed, surrounded by a supernova-remnant-like shell. These features all agree with the observations of 1E161348-5055. T{\.Z}Os are expected to have produced extraordinary high abundances of lithium and rapid proton process elements that would remain in the remnants and could be used to test this scenario.

\end{abstract}


\section{Introduction}

The X-ray source 1E161348-5055 (hereafter 1E1613) may be the most strange central compact object (CCO) in a supernova remnant (SNR). It has an extremely long periodic modulation, 6.67 hr \citep{De06}, while the surrounding SNR, RCW 103, is very young, $\sim$2000 yr \citep{Nugent,Carter}. It is located very close to the center of RCW 103, and has a variable X-ray luminosity in the range of $\sim 10^{33}-10^{35}$ erg s$^{-1}$, but has no identified radio, infrared, or optical counterpart. In addition, any other periodicities with $P\geq 12$ ms are excluded with high confidence \citep{De06}.

The origin of the 6.67 hr periodicity is puzzling. Some suggest that this source may be a low-mass binary with an underluminous secondary component which has an unusually low X-ray luminosity, and the orbital period is 6.67 hr \citep{Garmire00,De06}. However, it is difficult to explain how the close companion could have survived the supernova explosion and now escapes detection in multi-wavelength observations. In addition, one would also expect to detect the spin period of the compact star, assuming it to be a neutron star (NS). Some argue that 1E1613 may be a magnetar and was braked by a fall-back disk to such a long spin period \citep{De06,Li07}. This scheme is not in agreement with the observations of all the known magnetar candidates, anomalous X-ray pulsars, and soft gamma-ray repeaters among which a surrounding fall-back disk has been detected \citep{Wang06}, since the periods of these objects are all in the range of $2-12$ s, much shorter than 6.67 hr. In this work, we propose a new approach for the origin of the 6.67 hr period of 1E1613, namely, that it was the compact core of a Thorne-{\.Z}ytkow Object \citep{TZ75,TZ77}, in which it has been spun-down to its long period.

\bf {\em The Thorne-{\.Z}ytkow Object.} \rm
A Thorne-{\.Z}ytkow Object (hereafter T{\.Z}O) is a star that has a solar mass degenerate neutron core and a massive envelope, and looks like a red giant or supergiant \citep{TZ75,TZ77}. For its history and more information on T{\.Z}Os, see \cite{Landau32,Landau37,Gamow37,Eich89,Biehle,Cannon92}.
A T{\.Z}O cannot be formed from single star evolution. It is generally recognized that a T{\.Z}O is formed from the merger of a NS with a main-sequence star or a giant. There are three possible formation mechanisms: (1) If a newly born NS has an appropriate kick velocity, it could coalesce with its companion and immediately form a T{\.Z}O \citep{Benz92,Leonard94}; (2) The coalescence of a massive X-ray binary \citep{Taam78,Taam00}; (3) The capture of a NS by a main-sequence star in a globular cluster \citep{Ray87}. Based on the above mechanisms, \cite{Pod95} estimated a total number of 20-200 T{\.Z}Os existing in the Galaxy, given that the characteristic lifetime of a T{\.Z}O is $10^5-10^6$ yr.
The nuclear burning in a T{\.Z}O could produce extraordinary high abundances of lithium and rapid proton process elements \citep{TZ77,Cannon}, which makes it possible to distinguish a T{\.Z}O from an ordinary giant.

Some authors have argued that (at least massive) T{\.Z}Os cannot exist because neutrino-driven runaway accretion onto the NS in the spiral-in phase  will convert it into a black hole (see, e.g., \cite{Chevalier93, Brown95, Fryer99}). However, this idea seems difficult to reconcile with observations of the existence of tight-orbit double NSs (see the arguments given by \cite{Tauris06}). Also, a first T{\.Z}O candidate has now been found by \cite{Levesque14} in the Small Magellanic Cloud. The anomalous enhancements of the Rb/Ni, Li/K, Li/Ca, and Mo/Fe ratios in this candidate, HV 2112, satisfy the basic criteria for T{\.Z}O detection put forth by previous surveys and models (e.g., \cite{Biehle94,Pod95,Vanture99,Kuchner02}): so far, the only explanations for this chemical signature are the nucleosynthesis processes unique to the internal structure of a T{\.Z}O \citep{Levesque14}.

\section{The T{\.Z}O scenario for 1E1613 and RCW 103}

When a T{\.Z}O is just formed, its envelope usually spins very slowly because it is huge and has large moment of inertia, but the neutron core may still spin rapidly since it has been an NS \citep{Pod95}. In this section we will first investigate how the envelope can make the neutron core spin very slowly after a short time, and then explain why the core could become detectable. Finally, we present the possible T{\.Z}O scenario for 1E1613 and RCW 103.

\bf {\em Braking of the T{\.Z}O compact core.} \rm
The spin evolution of a neutron core in a T{\.Z}O has been discussed by \cite{Pod95}, who only considered the angular momentum carried by the accreted material. However, these authors did not take into account the coupling between the neutron core magnetic field and the envelope. The envelope is convective \citep{TZ77} and, of course, highly ionized. Therefore, the envelope plasma will be coupling with the rotating magnetic field.

In a T{\.Z}O envelope, the radial motion of the plasma is dominated by the gas and radiation pressure and gravity because both magnetic pressure and centrifugal force are relatively negligible. Nevertheless, the rotational motion of the plasma close to the neutron core is determined by the magnetic field. Close to the neutron core the magnetic field is sufficiently strong to make the plasma co-rotate with the neutron core. The radius up to which co-rotation exists is the Alfv\'en-radius $r_{\rm A}$. In this case, $r_{\rm A}$ is defined as the place where the magnetic energy density of the magnetic field $B^2/8\pi$ equals the kinetic energy density of the ram pressure experienced by the non-rotating plasma of density ¦Ñ at the magnetospheric boundary, due to this plasma being forced to co-rotate with the neutron star (e.g., \cite{Kundt76}), where $B$, $\rho$, and $\Omega$ are the magnetic field strength, envelope density, and core angular velocity, respectively.
One can approximate the envelope density profile around the neutron core as given by \cite{TZ77} with an exponential form, i.e., $\rho(r) = 10^{12}/r^2$ g cm$^{-3}$, and assume a dipole magnetic field, i.e., $B(r) = B_0(R_*/r)^3$, where $R_*$ and $B_0$ are the radius of the core and the field strength at the core surface. One then finds
\begin{equation}
r_{\rm A}^3 = B_0 R_*^3 / (10^6 (4\pi)^{0.5} \Omega).
\end{equation}
For typical parameter values, $B_0= 10^{12}$ G, $R_*=10^6$ cm, and $\Omega = 100 $ s$^{-1}$ ($P = 0.0628$ s), one finds $r_{\rm A} = 1.4\times10^7$ cm, which is $14R_*$, well outside the neutron core.

According to the work of \cite{Kundt76}, the swept-back field lines at the surface of the magnetosphere exert a braking torque on the magnetosphere of
\begin{equation}
T = dJ/dt \approx -0.5 B^2(r_{\rm A}) r_{\rm A}^3 = -0.5\times10^6 (4\pi)^{0.5} B_0 R_*^3 \Omega.
\end{equation}
(this is quite similar to the propeller torque introduced by \cite{Illarionov75}).
The angular momentum of the neutron core is
\begin{equation}
J=k^2 M R_*^2 \Omega,
\end{equation}
where $M$ is the mass of the neutron core and $k$ is its radius of gyration. Values of $k^2$ are typically of the order of $0.1$. Assuming this value, one obtains a spin-down timescale of the neutron core of:
\begin{equation}
t \approx M ln(\Omega_0) / (5\times10^6 (4\pi)^{0.5} B_0 R_*),
\end{equation}
where $\Omega_0$ is its initial spin angular velocity. We should mention here that the timescale given by Equation (4) is an upper limit to the real braking timescale, because when the value of $\Omega$ drops below about 10 s$^{-1}$, the ram pressure becomes lower than the thermal pressure at the magnetospheric boundary and from here on the (larger) thermal pressure will determine the (larger) braking torque. Assuming $M=3\times10^{33}$ g, $\Omega_0 = 100 $ s$^{-1}$, $B_0 = 10^{12}$ G and $R_* = 10^6$ cm, one finds with Equation (4) a spin-down timescale of about $t \sim 7.8\times10^8$ s, which is only about 25 yr.

The above picture requires some further clarification of how the angular momentum is transferred from the rotating NS to the convective envelope of the T{\.Z}O in our model. We envisage this to happen roughly as follows. At the magnetospheric boundary (Alfv\'en surface), the inflowing matter is forced to start co-rotating with the NS, which continues during its further infall. It can, of course, only flow in along the open field lines in two cones above the magnetic poles. As pictured in the \cite{Kundt76} model, there will be a transition layer at $r>r_A$ where the magnetic field is no longer strong enough to force the matter to co-rotate with the NS, such that there the inflowing matter sweeps the magnetic field lines backward. The resulting curvature of the field lines produces a strong magnetic tension which results in the above described braking torque. This torque implies that angular momentum  is carried off by matter near the magnetospheric boundary. We expect the magnetospheric boundary layer to be turbulent, such that a part of the matter is carried inward and gets attached to the field lines and accretes onto the neutron star, while another (smaller) part is carried off by the turbulence with its gained angular momentum, and is further carried outward by envelope convection. This matter can be considered to slip outward along swept-back magnetic field lines, maintaining its specific angular momentum, somewhat like to the solar wind matter coupled to the swept-back magnetic field lines anchored in the solar corona.

Before a T{\.Z}O was formed, when the NS was spiraling through the envelope toward the center of the companion star, the coupling of the magnetic field with the plasma could very likely already have occurred. At the same time, the matter accreted onto the star surface will exert an opposite torque. The torque of peaceful accretion is much smaller than that of the magnetic coupling; thus, the total torque should already brake the rotation of the compact star during the spiral-in phase. Considering, however, that the spiral-in timescale is very short (see below) and the subsequent braking is already extremely fast, we ignore the spin evolution during this spiral-in phase.

\bf {\em Destruction of the T{\.Z}O envelope.} \rm
The final fate of a T{\.Z}O is an open question. Here we discuss the possibility that the envelope may be exhausted by its strong stellar wind or be disrupted by some powerful bursts before the neutron core transforms into a black hole. The neutron core could then be detectable as a very slowly spinning CCO in an SNR-like shell.

The T{\.Z}O envelope has a low density and a very large radius, which means that its gravitational binding energy is not large. This loosely bound envelope can therefore be easily lost. We estimate upper and lower limits to the binding energy of the envelope, by numerically integrating the mass distributions given in Tables 1 and 2 of \cite{TZ77} for $5 M_\odot$ and $12 M_\odot$ T{\.Z}O stars, respectively. Since the density decreases outward, one obtains an upper limit to the binding energy of the envelope if one assumes that the density at the inner edge of a spherical shell is the mean density of this shell, and then integrates over all shells. Likewise, by assuming the density at the outer edge of the shell to be the mean density of the shell, one obtains by integration over all shells a lower limit to the binding energy of the envelope. From these integrations, one finds for the $5 M_\odot$ T{\.Z}O that the binding energy of the envelope is in the range $(2.1-3.6) \times 10^{46}$ erg and for the $12 M_\odot$ T{\.Z}O it is in the range $(4.9-14.2) \times 10^{46}$ erg.

These values are quite low, which means that the envelopes can be very easily lost, either by a strong stellar wind of the T{\.Z}O or by a burst of energy larger than the binding energy of the envelope, produced for some reason by the CCO. In the following, we assume, for the sake of argument, a binding energy of the envelope of $10^{47}$ erg. A T{\.Z}O usually has high luminosity, which means that its stellar wind could be extremely strong: red supergiants  with a similar radius and luminosity typically have mass-loss rates of about $10^{-5}M_\odot$ yr$^{-1}$, such that a $10 M_\odot$ envelope is  lost in $\sim 10^6$ yr. However, as pointed out by \cite{TZ77}, T{\.Z}O envelopes with a mass below about one $M_\odot$ become dynamically unstable (similar to Mira variables) with a mass-loss rate $\sim 10^{-3}M_\odot$ yr$^{-1}$, and therefore one may expect that once the envelope mass has been reduced to one $M_\odot$, this envelope will be ejected in a very short time, of about $10^3$ yr. This could have produced the shell around 1E1613.

An alternative of the production of this shell by this superwind is that two kinds of bursts may have been produced by the compact stellar core that may have led to the ejection of the envelope: (1) \cite{Ouyed} proposed that an NS could transform into a quark star, which releases an extremely large amount of energy in a so-called quark-nova. If this happens after the neutron core has been spun down, the envelope can be disrupted immediately. (2) If the compact core is born as a quark star, then when the star has accreted enough material to have formed a hadron crust, the hadron crust may break and transform into quark matter \citep{Cheng}. The energy efficiency of the phase transition is of the order of magnitude $10\%$. If such a crust phase transition occurs, the energy released can easily be larger than the binding energy of the envelope, and could destroy this envelope immediately.

After the envelope is disrupted, the core will be detectable with an extremely long spin period. The destroyed envelope could form an SNR-like shell.

On the basis of the above described possibilities for the T{\.Z}O envelope ejection, the following two T{\.Z}O scenarios for the formation of 1E1613 and RCW 103 can be envisaged.

Case A. About 2000 yr ago, the more massive star of a binary experienced a supernova explosion, which produced RCW 103 and a compact star. The compact star received an appropriate kick velocity, which caused it to directly coalesce with the companion star and form a T{\.Z}O \citep{Benz92,Leonard94}. Then, the spin of the compact core was braked  quickly and the envelope -- if it had a low mass -- was exhausted by a strong stellar wind or was destroyed by some powerful burst within 2000 yr, as described above. As a result, the envelope mixed with the original SNR RCW 103 and the core became 1E1613.

Case B. More than 2000 yr ago (up to one million years ago) a T{\.Z}O formed due to spiral-in of a high-mass X-ray binary. Then the spin of the  core was quickly braked to a period of several hours. About 2000 yr ago, either (1) a powerful burst (quark-nova or crust phase transition) occurred which destroyed the envelope and turned it into a supernova-like shell: RCW 103, and the quark core became 1E1613; or (2) in $\sim10^6$ yr the envelope mass was reduced to one $M_\odot$, after which it became dynamically unstable and was ejected in $\sim$1000 yr, forming the shell RCW 103.

In both of these cases, the present X-ray emission of the compact star could either be due to the fact that the compact star is still very hot (due to its recent accretion history), or due to accretion from a residual envelope or from a fall-back disk.

In case A, all the events should have happened within about 2000 yr; thus, the timescales are very critical. If one assumes the companion mass, the orbital semimajor radius, and the kick velocity to be $5 M_{\odot}$, $10^{15}$ cm, and 300 km s$^{-1}$, then the NS could touch the companion about one year after its birth and spiral into the center within several months (this timescale is difficult to calculate exactly but can be roughly estimated \citep{Benz92,Taam00}). When the NS arrived in the center of the companion, nuclear burning will have ignited after a free-fall timescale of several months, after which it became a T{\.Z}O. All the above timescales are much shorter than 2000 yr, the spin-down timescale usually is shorter than 2000 yr, and the envelope could also be destroyed within 2000 yr. Consequently, the timescales of case A can match with the age of RCW 103. In case B, the timescale is not a problem, but in Case B (1) a phase transition process is necessary, which may imply that 1E1613 is a quark star.

If 1E1613 has undergone a T{\.Z}O phase, it should have a very slow proper motion, since the massive companion absorbed most of the kick impulse of the NS. This is consistent with the observation of 1E1613. In the T{\.Z}O scenario, SNR RCW 103 should be different from a real SNR, since the lithium and rapid proton process elements produced during the T{\.Z}O phase should remain in RCW 103, which could be measured in the future and serve as a test of the scenario.

\section{Discussion}

T{\.Z}Os should exist in our Galaxy since an NS has various possibilities to enter into a normal star \citep{Taam78,Ray87,Benz92,Taam00}.
In this Letter, we studied the spin evolution of the magnetized compact core inside a T{\.Z}O, and found that it could evolve into a extremely long-period isolated X-ray source, surrounded by an SNR-like shell. The period and proper motion of 1E1613 and the age of RCW 103 are consistent with the T{\.Z}O scenario. This scenario could be tested by measuring the elemental abundance of RCW 103, which may be very difficult, especially for case A whose T{\.Z}O life time is only about 2000 yr, which means that the production of rp-process elements is probably quite limited.

The T{\.Z}O scenario has two advantages for producing an extremely long period compact star: (1) the strong coupling between the envelope and the magnetic field can brake the core very quickly; (2) the very slowly spinning envelope can brake the core to an extremely long spin period. Comparatively, braking by a stellar wind usually needs a very long timescale, and braking by a disk produces an equilibrium period that usually is not very long.

From the birth rate of T{\.Z}Os, $\sim2\times10^{-4}$ yr$^{-1}$ \citep{Pod95}, if we assume one-fourth of them to evolve to a long-period compact star, of the order of $10^5$ of such stars would have formed in the Galaxy. Of course, most of them should be undetectable because they are old, cool, isolated, and slowly rotating. However, if they are accreting from the residual envelope and have a large luminosity, they could be detected, such as 1E1613. If future observations confirm this scenario, it would indirectly prove the existence of T{\.Z}Os and give strong constraints on common-envelope evolution \citep{Ivanova13}. The T{\.Z}O envelope could be destroyed within $\sim1000$ yr, as shown in Section 2, which is much shorter than the estimated T{\.Z}O lifetime of $10^5-10^6$ yr \citep{Pod95}. Therefore, the expected number of T{\.Z}Os in the Galaxy could be smaller than one, which can explain why none have been detected so far in our Galaxy.

\acknowledgments

We thank the referee and the pulsar group of PKU for helpful discussions, comments, and suggestions.
This work is supported by the National Basic Research Program of China (grant No. 2012CB821800), the National Natural Science Foundation of China (grant Nos. 11225314, 11033008, 11373011), and the XTP project XDA04060604.

\clearpage

\end{document}